# Effects of particle shape and size distribution on hydraulic properties of grain packs: An experimental study

Elnaz Rezaei [a], Kamran Zeinalzadeh [a, *], Behzad Ghanbarian [b]

[a]Department of Water Engineering, Faculty of Agriculture, Urmia University, Urmia, 57135-165, Iran

[b]Porous Media Research Lab, Department of Geology, Kansas State University, Manhattan 66506 KS, USA

*Corresponding author: Email: k.zeinalzadeh@urmia.ac.ir   Tel/Fax: +9844-32775035

## Abstract

Uniform and multi-dispersed grain packs have been frequently used to conceptually study flow in porous media. Numerical simulations were previously used to address the effect of particle shape on characteristics, such as pore space fractal dimension, moisture characteristic curve (MCC) and saturated hydraulic conductivity (SHC) of grain packs. However, experimental observations are still required since fractal-based approaches have been extensively proposed to model various properties in porous media. In this study, 16 angular sand and 16 spherical glass bead samples with different particle size distributions (PSDs) from well- to poorly-sorted were packed. The MCC was measured using the combination of sandbox and pressure plates methods. The pore space fractal dimension ($D_{MCC}$), calculated from the measured MCC, ranged from 0.80 to 2.86 in sand and from -0.18 to 2.81 in glass bead packs, which indicated that $D_{MCC}$ may be negative in homogeneous media (e.g., glass bead packs) consistent with several studies in the literature. Results showed greater $D_{MCC}$ for the sand packs than the glass bead packs with the same geometric mean diameter values and PSDs. This clearly demonstrated the effect of particle shape on $D_{MCC}$ in the studied packs. The critical path analysis (CPA) approach was used to estimate the SHC measured using the constant-head method. We found that the CPA estimated the SHC accurately, within a factor of four of the measurements on average. Although the CPA is theoretically known to be accurate in media with broad pore size distributions, we experimentally found that it estimated the SHC in various types of grain packs reasonably well.

## 1. Introduction

Moisture characteristic curve (MCC) and saturated hydraulic conductivity (SHC) are among the important hydraulic properties of porous media. These parameters are widely used to model water movement and solute transport under variably-saturated conditions. However, their direct measurements are costly and time-consuming, and, therefore, their modeling and indirect estimation have been the subject of active research.

MCC models are typically classified into two groups: (1) empirical and (2) theoretical. In the latter, fractal geometry (Mandelbrot, 1982) has been frequently applied to relate the MCC to pore and particle size distributions (Li and Horne, 2009; Soldi et al., 2017; Hunt et al., 2014). One basic advantage of theoretical approaches and fractal models is that their parameters are physically meaningful. In addition to that, pore and particle characteristics can be linked through mathematical relationships (see, e.g., Bird et al., 2000; Ghanbarian-Alavijeh and Hunt, 2012a). Various fractal models have been developed to estimate MCC based on self-similar properties of soils and rocks. Avnir et al. (1998) stated that, "A fractal object, in the purely mathematical sense, requires infinitely many orders of magnitude of power-law scaling, and a consequent interpretation of experimental results as indicating fractality requires many orders of magnitude". Crawford et al. (1995) are among the first who demonstrated that the fractal dimension derived from moisture characteristic curve measurements was near that determined from images of soils. Similar results later were obtained by Dathe and Thullner (2005). Although mathematically a fractal object needs to span infinite orders of magnitude of scale, a prefractal object has a finite number of iterations and thus is limited in range (Ghanbarian and Millán, 2017). Sahimi (2003) also stated that natural porous media like soils and rocks that show self-similarity typically lose their fractal properties below the lower cutoff and above the upper cutoff scales.

In addition to the MCC, the SHC is another quantity affected by microscopic properties of pore space, such as pore connectivity and pore size distribution. Numerous studies have modified the Kozeny-Carman model to estimate the SHC. For instance, Xu and Yu (2008) derived an analytical expression for the SHC using fractal properties of porous media and a bundle of tortuous tubes. Their study provided a theoretical expression for the constant coefficient in the Kozeny-Carman model. However, the Kozeny-Carman model is of approximate validity. It is often invalid in the case of grains deviating strongly from spherical shape, broad grain size distributions, and consolidated materials. Therefore, it should be used with great caution (Dullien, 1992).

Alternative models based on concepts from critical path analysis (CPA) and percolation theory (Katz and Thompson, 1986), effective-medium approximations (Doyen, 1988), self-consistent models (Dagan et al., 2013), and wavelet transforms (Moslehi et al., 2016) were accordingly developed to estimate the SHC. For example, Katz and Thompson (1986) used the CPA approach to estimate permeability of rocks. Based on CPA and percolation theory, one can define a critical pore size below which the contribution of pore conductance to macroscopic flow and permeability is trivial. In the Katz and Thompson (1986) model, permeability is estimated from the formation factor and critical pore diameter. The latter was determined from the inflection point of mercury intrusion capillary pressure curve. Their results showed that permeability was estimated within a factor of two of measured values in sandstones and carbonates.

Recently, Ghanbarian et al. (2017b) developed a new theoretical model based on concepts of CPA to estimate SHC in soils from measured moisture characteristic curves. They reported SHC estimations to be within a factor of 3 of the average measured SHC reported by Rawls et al. (1982) for eleven USDA soil texture classes. Those authors argued that the SHC overestimation might be due to underestimating the formation factor or overestimating critical pore radius. The CPA estimated the SHC of fine-textured soils more accurately than that of coarse-textured samples (see their Fig. 4), which indicated that CPA might be a more appropriate upscaling method for porous media with broad pore throat-size distributions.

Although the effect of particle shape on SHC has been addressed in the literature (see, e.g., Sperry and Peirce, 1995; Garcia et al., 2009; Torskaya et al., 2014), its influence on pore space fractal dimension has remained as an open question. In addition to that, recently, negative pore space fractal dimensions were reported for numerically simulated MCCs in mono-sized sphere packs (see, e.g., Ghanbarian and Sahimi, 2017), while its value typically ranges between 0 and 3 in natural porous media like rocks and soils. Although Ghanbarian-Alavijeh and Hunt (2012b) theoretically demonstrated that the pore space fractal dimension ranges between $-\infty$ and 3, to the best of the authors' knowledge, there exists no experimental evidence for negative fractal dimension values in the hydrology and soil science communities. To address this knowledge gap, it is necessary to assess how the pore space fractal dimension varies with the particle-size distribution broadness in homogeneous and relatively heterogeneous media. Therefore, the main objectives of this *experimental* study are to: (1) investigate the effect of particle shape on the pore space fractal dimension, (2) determine how the pore space fractal dimension varies with the particle-size distribution broadness in homogeneous and relatively heterogeneous packs, and (3) demonstrate that pore space fractal dimension may take negative values (4) evaluate the CPA model and its predictability for estimating the SHC in uniform media with narrow particle size distributions.

## 2. Materials and Methods

### 2.1. Sample Preparation

In this study, we used two particle types, i.e., angular sands and spherical glass beads, to generate synthetic packs of various heterogeneities. The sand particles were riverbed sediments collected from the Nazlou River, Urmia Lake Basin in Iran, while the glass beads were smooth spheres with a high degree of sphericity (Fig. 1). Both sand and glass bead particles were classified using the wet-sieve method and sieve numbers 12, 14, 16, 25, 35, 50, 60, 80, 100, 120, and 140 meshes (Nimmo and Perkins, 2002) and then washed with deionized water until the drained solution electrical conductivity reached to 5 μs/cm or less.

Fig. 1

We generated sixteen sand and sixteen glass bead samples by packing the particles within plexiglass columns with 4.4 cm internal diameter and 5 cm height. The dry packing method was applied as the most common approach to produce homogeneous columns (Oliviera et al., 1996; O'Carroll et al., 2004; Plummer et al., 2004; ASTM:D421−85, 2007; Lewis and Sjöstrom, 2010). Dry particles were added to the plexiglass column by depositing 0.2 cm layers (Oliviera et al., 1996), compated using a wooden plunger, and then vibrated. To reach the top of the column, the procedure was replicated several times.

For each sample, bulk and particle density values ($\rho_b$, $\rho_s$) were determined, and porosity was calculated from such measurements ($\phi = 1 - \rho_b/\rho_s$). To achieve a wide range of classes from uniform to relatively heterogeneous, sand and glass bead particles were packed within four classes from A (well-sorted) to D (poorly-sorted). Class A includes seven uniform packs (e.g., subclasses A1 to A7) with particles sorted between two sieve sizes listed in Table 1. Geometric mean diameter (GMD $= \exp\left(\frac{\sum_{i=1}^{n} w_i \ln d_i}{\sum_{i=1}^{n} w_i}\right)$ in which $w_i$ is the weighting factor, and $d_i$ is the particle diameter) was determined for each sample to represent the average particle size (Porter et al., 2013). As can be seen from Table 1, the GMD decreases from 0.65 mm for subclass A1 to 0.12 mm for subclass A7. In class A, the value of porosity is about 0.41 for sand packs, while its value ranges between 0.37 and 0.42 for glass bead packs. Three different particle sizes from class A were equally mixed to generate class B and prepare four subclasses named B1 to B4 (see Table 1). Within class B, the largest and smallest GMD values belong to subclasses B1 and B2, respectively. The porosity varies from 0.36 (B2) to 0.42 (B1) for sand and from 0.35 (B2) to 0.41 (B4) for glass bead packs. Samples in class C were packed by mixing five different particle sizes from class A with equal weights. The average porosity value within class C is 0.405, and GMD ranges between 0.18 mm (C2) to 0.34 mm (C1). Class D, the most heterogeneous packing, consists of only one sample, including seven particle sizes from class A. The value of GMD for sample D is 0.25 mm, and the porosity is 0.4 for both sand and glass bead packs. The salient properties of the sand and glass bead packs are summarized in Table 1. We should point out that to ensure the uniformity of the samples and increase the accuracy of the measurements, the same packs were used for all the measurements, including the MCC and SHC, under constant room-temperature conditions.

Table 1

## 2.2. MCC and SHC Measurements

To measure the MCC a combination of the sandbox and pressure plate methods (ASTM:D6836-16, 2016) was used under the drainage conditions. The water content ($\theta$) was determined at the tension heads 0, 1, 2.5, 10, 15, 31.6, 48, 63.1, 80, and 100 cm of $H_2O$ using the sandbox apparatus (Eijkelkamp Agrisearch Equipment, Giesbeek, The Netherlands) and by hanging water column (Stakman et al., 1969). For the sake of measurement accuracy, well- and poorly-sorted samples were separately analyzed in the sandbox. To prevent evaporation from the sample surface, the sandbox lid was kept closed during the experiment. The MCC measurements were continued at higher tension heads e.g., 0.18, 0.2, 0.22, 0.24, 0.26, 0.28, 0.3, 0.34, 0.44, 0.48, 0.7, 1.2, 2, 3, 4.9, 7, 10, 12.6 and 15 bars using the pressure plate apparatus by adjusting the compressed gas cylinder–pressure regulator under equilibration conditions (Cresswell et al., 2008). Such measurements were performed only for samples SA6, SA7, SB4, SC3, GA5, GA6, GB2, GB3, GB4, GC3, and GD that retained a volumetric water content greater than 8% at the tension head of 100 cm of $H_2O$ in the sandbox apparatus. In both sandbox and pressure plate, all samples were first fully saturated (zero tension head conditions) using distilled water by allowing water to wick upward through the column (Lewis and Sjöstrom, 2010). The MCC measurement was ceased when the difference between the last two measured volumetric water content was less than 0.2%. The gravimetric water content was determined by weighing (±0.001 g) each sample at different tension heads and after drying at 105 °C for 24 hrs. The volumetric soil water content ($cm^3/cm^3$) was accordingly calculated using the bulk density.

The SHC values were determined using the constant head laboratory method (Klute and Dirksen, 1986) after the MCC measurements. Water flow was established from the bottom to the top of the columns to guarantee total saturation and minimize the entrapped air effects. The inlet and outlet boundaries of each sample were kept constant during the SHC experiments. After reaching the steady-state conditions, the SHC was calculated using Darcy's law (Jačka et al., 2014).

### 2.3. Pore Space Fractal Dimension Calculation

In order to calculate the pore space fractal dimension (hereafter $D_{MCC}$), the following fractal moisture characteristic curve model (Ghanbarian et al., 2017a) was fit to the measured MCCs using the Curve Fitting toolbox of MATLAB and the Levenberg-Marquardt algorithm:

$$S_w = 1 - \frac{\beta}{\phi}\left[1 - \left(\frac{h}{h_{min}}\right)^{D_{MCC}-3}\right] \quad (1)$$

In Eq. (1), h is the suction (or tension) head, $h_{min}$ is the air entry pressure, $S_w$ is the water saturation ($= \theta/\phi$), and $\beta = (\phi - \theta_r)r_{max}^{3-D_{MCC}}/(r_{max}^{3-D_{MCC}} - r_{min}^{3-D_{MCC}})$ in which $\theta_r$ is the residual water content, and $r_{min}$ and $r_{max}$ ($= 0.149/\ h_{min}$ in which $h_{min}$ is in cm $H_2O$) are the smallest and largest pore radii in the medium, respectively.

Eq. (1), mathematically derived from the truncated power-law pore size distribution (i.e., $f(r) \propto r^{-1-D_{MCC}}$), reduces to the Brooks–Corey model (Brooks and Corey, 1964) as its special case. However, it is not an explicit function of $\theta_r$ (or residual water saturation, $S_{wr}$). Another form of Eq. (1) is (Ghanbarian et al., 2017a):

$$\frac{S_w - S_{wr}}{1 - S_{wr}} = \frac{h^{D_{MCC}-3} - h_{max}^{D_{MCC}-3}}{h_{min}^{D_{MCC}-3} - h_{max}^{D_{MCC}-3}} \quad (2)$$

through which one can estimate the value of residual water saturation ($S_{wr}$) by directly fitting Eq. (2) to the MCC measurements, as conducted in this study.

### 2.4. SHC Estimation

Within the CPA framework, flow is mainly controlled by pores with sizes greater than a critical value ($r_c$). Using concepts of CPA and fractal properties of the pore space, Ghanbarian et al.

(2017b) derived a theoretical model to link the critical pore radius $r_c$ to other properties of the porous medium e.g., critical water content $\theta_t$, $\beta$ and $r_{max}$. Such a relationship i.e., $r_c = r_{max}\left[1 - \frac{\theta_t}{\beta}\right]^{\frac{1}{3-D_{MCC}}}$ in combination with the Katz and Thompson (1986) model was proposed to estimate the SHC as follows (Ghanbarian et al., 2017b):

$$\text{SHC} = f_f \frac{A^2 h_{min}^{-2}}{C_{KT}} \frac{\phi}{3}\left[1 - \left(\frac{1}{3}\right)^{3-D_{MCC}}\left(1 - \frac{\theta_t}{\beta}\right)\right]\left(1 - \frac{\theta_t}{\beta}\right)^{\frac{2}{3-D_{MCC}}} \qquad (3)$$

In Eq. (3), SHC represents the saturated hydraulic conductivity, $\phi$ is the porosity, $f_f$ is the fluidity factor (= $\rho_f g/\mu$, $\rho_f$ is the fluid density, g is the gravitational acceleration and $\mu$ is the dynamic fluid viscosity). The constant A is the coefficient in the Young-Laplace equation (=$2\gamma\cos(\omega)$ in which $\gamma$ is the air-water interfacial tension and $\omega$ is the air-water contact angle), and $C_{KT}$ is the Katz-Thompson constant coefficient and equal to 56.5. Note that $h_{min}$, $\beta$ and $D_{MCC}$ are optimized by fitting Eq. (1) to the MCC measurements. The critical volume fraction for percolation ($\theta_t$) is not a priori known in soils, although it can be roughly estimated from the value of residual water content ($\theta_t \approx \theta_r$) (Hunt et al., 2014).

In addition to the CPA approach, we estimated the SHC using the Kozeny-Carman model (Carman, 1937; Kozeny, 1927), a widely used model that is typically presented in the following form:

$$\text{SHC} = f_f \frac{\bar{D}^2 \phi^3}{180(1-\phi)^2} \qquad (4)$$

where $\bar{D}$ is the representative grain diameter. Porter et al. (2013) recommended that the representative grain diameter should be calculated using the geometric mean, if unoccupied coarse pores percolate. Otherwise, the harmonic mean should be used to determine $\bar{D}$. For the sake of simplicity, we used the GMD values as representative diameter $\bar{D}$ to estimate the SHC via the Kozeny-Carman equation.

### 2.5. Evaluation Criteria

To statistically assess the accuracy of Eqs. (1) to (4), we calculated the correlation coefficient ($R^2$), the root mean square error (RMSE) and the root mean square log-transformed error (RMSLE):

$$R^2 = \frac{n(\sum_{i=1}^{n} P_F^{(i)} P_M^{(i)}) - (\sum_{i=1}^{n} P_F^{(i)} \sum_{i=1}^{n} P_M^{(i)})}{\sqrt{[n\sum_{i=1}^{n} P_F^{(i)2} - (\sum_{i=1}^{n} P_F^{(i)})^2][n\sum_{i=1}^{n} P_M^{(i)2} - (\sum_{i=1}^{n} P_M^{(i)})^2]}} \quad (5)$$

$$RMSE = \sqrt{\frac{\sum_{i=1}^{n}(P_M^{(i)} - P_F^{(i)})^2}{n}} \quad (6)$$

$$RMSLE = \sqrt{\frac{\sum_{i=1}^{n}(LogP_M^{(i)} - LogP_F^{(i)})^2}{n}} \quad (7)$$

$P_F^{(i)}$, $P_M^{(i)}$ and n are the fitted values, the measured values, and the number of the data points, respectively. The smaller the RMSE and RMSLE values, the higher the accuracy.

We also plotted the Taylor diagram (Taylor, 2001) that shows the correlation coefficient $R^2$ against the standard deviation. It indicates the degree of match between the measured and the fitted water content values through either Eq. (1) or (2). In this diagram, as $R^2$ tends to 1, the standard deviation approaches zero, thus the higher the accuracy. We also compared the values of $D_{MCC}$ and SHC in sand packs with those in glass bead packs using the paired t-test method at the 95% confidence level.

### 3. Results

To study the effect of particle shape and PSD on the $D_{MCC}$, Eq. (1) was fit to the measured MCCs for all sand and glass bead packs. The optimized parameters $D_{MCC}$, $\beta$ and $h_{min}$, their values as well as the RMSE and $R^2$ values are given in Table 2 for each sample. In the following, we present results for various packs from class A to D. The SHC values are also reported for both sand and glass bead packs.

Table 2

### 3.1. $D_{MCC}$ and MCC Variations

**- Class A**

Results presented in Table 2 show that $D_{MCC}$ ranged between 0.80 and 1.80 for the SA samples (including SA1 to SA7) and between -0.18 and 1.39 for the GA samples (including GA1 to GA7). For sample GA3 with geometric mean diameter of 0.32 mm, we found a negative pore space fractal dimension ($D_{MCC}$ = -0.18). For all samples in class A, the value of the $D_{MCC}$ is less than 2.0 corresponding to a relatively homogeneous pore structure. Fig. 2 demonstrates the $D_{MCC}$ variation with the GMD for sand and glass bead packs in class A. As can be seen, there is no clear trend between the $D_{MCC}$ and the particle GMD. Although the GMD value monotonically decreases from A1 (coarse-textured, (NRCS, 1993)) to A7 (very fine-textured, (NRCS, 1993)),

Fig. 2

Fig. 3 presents the measured MCC as well as the fitted model, Eq. (1), for fourteen samples in class A (including both sand and glass bead packs). The value of residual water content – determined from directly fitting the fractal MCC model, Eq. 2, to the measured data – varied from 4 to 7% within SA samples (SA1 to SA7) and from 1 to 10% within GA samples (GA1 to GA7).

Fig. 3

Generally speaking, the fine-textured samples (e.g., A4 to A7 and G4 to G7 with 0.125 < GMD <0.25 mm) exhibited greater pore space fractal dimension and accordingly more complex pore structure (see Fig. 2). The calculated RMSE values were less than 0.03 and 0.01 (cm$^3$/cm$^3$) for sand and glass bead packs, respectively (Table 2).

**- Class B**

In class B, we found the pore space fractal dimension varied between 1.87 and 2.02 for sand and between 1.42 and 2.07 for glass bead packs. $D_{MCC}$ variation with GMD is shown in Fig. 4 for samples in class B. As can be seen, there is no clear relationship between the pore space fractal dimension and the geometric mean diameter (Fig. 4).

Fig. 4

Fig. 5 shows the measured MCCs as well as the fitted model, Eq. (1), for samples in class B. Although the measured MCCs are very similar for SB1 and GB1, those curves are relatively dissimilar for other samples. The optimized residual water content for the sand and glass bead packs varied from 4 to 8% and from 3 to 15%, respectively, which are not greatly different from similar samples in class A. For samples in class B, the RMSE values for MCC estimation ranged from 0 to 0.01 and 0.01 to 0.020, for sand and glass bead packs, respectively (Table 2).

Fig. 5

**- Classes C and D**

The pore space fractal dimension varied in the range of 1.89 to 2.86 for sand and 1.66 to 2.81 for glass bead packs in class C (Table 2). We found $D_{MCC}$ = 1.75 and 1.81 for samples SD and GD, while $D_{MCC}$ = 2.86 and 2.57 for samples SC4 and GC4, respectively (Fig. 6). The optimized residual water content ranged between 6 and 8% in the sand and between 8 and 15% in the glass bead packs.

Fig. 6

Fig. 7 shows the measured MCCs as well as the fitted model, Eq. (1), for samples in classes C and D. The RMSE values obtained from fitting Eq. (1) to the measured MCCs varied in the range of 0 to 0.02 (cm$^3$/cm$^3$) for sand packs (e.g., SC and SD) and from 0 to 0.03 (cm$^3$/cm$^3$) for glass bead packs (e.g., GC and GD) in classes C and D (Table 2). Results showed that Eq. (1) slightly better represented the MCCs in class B than those in classes C and D.

Fig. 7

**-Taylor Diagram**

Fig. 8 shows the Taylor diagram including the correlation coefficient $R^2$, determined from fitting Eq. (1) to the measured MCCs, against the standard deviation between the measured and fitted volumetric water contents. The correlation coefficient for the sand samples was found to be higher than 0.98, except for sample A7 ($R^2 = 0.96$). We found $R^2 > 0.95$ for glass bead packs. As can be seen in Fig. 8, standard deviation ranged between 0.04 and 0.13 for all samples including sand and glass bead packs. Results presented in Fig. 8 indicate that Eq. (1) was accurately fit the MCC measurements.

Fig. 8

**3.2. SHC Variations**

Fig. 9 shows the correlation between the SHC and the GMD for both sand and glass bead packs. As the GMD increases, the SHC increases as well. Interestingly, by only including those samples in class A we found a highly correlated power-law relationship between the SHC and GMD. More specifically, we found exponents near 2.00 and 3.53 with $R^2 = 0.93$ and 0.91 for the sand and glass bead packs, respectively.

Fig. 9

Fig. 10 shows the measured SHC as a function of the calculated critical pore radius (Table 2). The exponents 1.81 and 1.91 (with $R^2$ = 0.87 and 0.92) respectively for the sand and glass bead packs are not greatly different from the theoretical value of 2, derived by Katz and Thompson (1986). Comparing Figs. 9 and 10 clearly demonstrates that the SHC is more correlated to the critical pore radius than the GMD in the sand and glass bead packs.

Fig. 10

## 4. Discussion

Accurate characterizations of soil hydraulic parameters, such as MCC and SHC, under variably saturated conditions are essential for various environmental studies. It is well documented in the literature that fractal dimensions can be successfully used to estimate soil hydraulic properties (Hunt et al., 2014). In what follows, we discuss the effect of PSD on the MCC and SHC in the sand and glass bead packs studied here. The accuracy of the critical path analysis approach and Kozeny-Carman model (Carman, 1937; Kozeny, 1927) in the estimation of the SHC is also analyzed.

### 4.1. Effect of Particle-Size Distribution on $D_{MCC}$

For all subclasses except B3, C2, C3, and D, we found that the $D_{MCC}$ values in the sand packs were greater than those in the glass bead packs. By means of image analysis and from area and perimeter measurements, Arasan et al., (2011) reported the boundary fractal dimension of particles of various sphericities. Those authors found that the boundary fractal dimension increased as angularity increased or roundness decreased. Similar results were obtained by Araujo et al. (2017) who reported that sand particles with greater roughness had greater surface area and boundary fractal dimension.

The $D_{MCC}$ value does not follow any trend with the texture in the SA samples (Fig. 2). This is in accord with the results of Ghanbarian et al. (2018) who reported $D_{MCC}$ values for four sand and one glass bead packs. Similarly, those authors did not observe any relationship between $D_{MCC}$ and the particle size in relatively homogenous packs. However, it may not be necessarily valid in natural soils and rocks with broad PSDs and pore size distributions where there exist particles of different shapes and geometries. Interestingly, Ghanbarian et al. (2018) found $D_{MCC}$ = 1.32 for Granusil sand pack with average particle size 0.5 mm and $D_{MCC}$ = 0.98 for glass bead pack (see their Table 1). These values are in well agreement with the $D_{MCC}$ values found in this study for samples SA2 and GA2 ($D_{MCC}$ = 1.25 and 1.0, respectively). More specifically, Ghanbarian et al. (2018) noticed that the $D_{MCC}$ ranged between 0.98 for 0.5 mm glass bead pack and 1.76 for 0.3 mm Accusand pack with porosity near 0.4. Their results showed that samples with broader pore size distributions had larger pore space fractal dimensions. We should, however, point out that $D_{MCC}$ in natural porous media like soils is typically expected to vary between 2 and 3. For instance, Ghanbarian et al. (2016) reported the value of $D_{MCC}$ in the range of 2.38 to 2.97 for soil samples from the UNSODA database.

In both sand and glass bead packs, the $D_{MCC}$ for subclass A3 was lower than that for subclasses A1 and A2. Interestingly, we found a negative value for the pore space fractal dimension ($D_{MCC}$ = -0.18) for sample GA3 with geometric mean diameter of 0.32 mm. There exist theoretical (Ghanbarian-Alavijeh and Hunt, 2012b) and numerical (Ghanbarian and Sahimi, 2017) evidence that the pore space fractal dimension can take negative values. Negative fractal dimensions were also supported by Mandelbrot (1990). In the power-law probability density function of pores (i.e., $f(r) \propto r^{-1-D_{MCC}}$) the exponent ($-1- D_{MCC}$) controls the heterogeneity of pores and their size distribution. When $-1 < D_{MCC} < 3$, smaller pores are more probable than the larger ones. For $D_{MCC}$

= −1, the power-law probability density function reduces to the uniform distribution. However, when $D_{MCC} < -1$, larger pores are more probable than the smaller ones (Ghanbarian and Sahimi, 2017).

High residual water content values could be due to an increase in the particle effective surface irregularity for the samples in class A. To investigate the effect of surface roughness on residual water content, Dullien et al. (1989) measured drainage and imbibition capillary pressure curves in packings of untreated (smooth) and etched (rough) glass beads with respectively average size of 0.55 and 0.63 mm (see their Table 1). They observed a definite irreducible saturation in smooth glass bead packs, while its value kept decreasing as capillary pressure increased in etched glass bead packs. Dullien et al. (1989) argued that it was the surface roughness permitted much lower values of residual saturation in the etched glass bead packs.

Comparing the $D_{MCC}$ values in class B with those in class A showed that the pore structure of the samples in class B is more complex and heterogeneous than that in class A, as expected. However, we found that the pore space fractal dimension varied within a narrower range for the samples in class B with finer texture compared to those in class A.

We found the largest $D_{MCC}$ values for the samples SC4 and GC2 ($D_{MCC}$ = 2.86 and 2.81, respectively; Table 2). According to Ghanbarian and Millán (2017), greater fractal dimensions (such as SC4 and GC2 samples) are typically associated with broader pore and/or particle size distributions. We also found that the residual water content in the glass bead packs composed of smooth particles was greater than that in the sand packs constructed of angular grains only for three samples of class C.

Rokhforouz and Amiri (2019) investigated the effect of grain shape and size via numerical simulations. In their study, grain shapes were simplified by smooth spheres or approximated by irregular particles mimicked real grain shapes. For both cases, Rokhforouz and Amiri (2019) studied water propagation under neutral wetting conditions (i.e., 90° contact angle). They observed a thick water finger with less regularity shape and thickness in the real-shaped model with capillary pressure increment. This means that angular sands with rough surfaces could have a lower residual water content compared to smooth glass beads, consistent with the results of Dullien et al. (1989).

Our results showed that the $D_{MCC}$ values for the sand packs were highly correlated with that for the glass bead packs. In all classes except class A, the paired t-test results showed no significant difference between the $D_{MCC}$ values for the sand and glass bead packs at the 95% confidence level. More specifically, we found $(D_{MCC})_G = 1.326(D_{MCC})_S - 0.79$, in which the indices G and S stand respectively for glass bead and sand, with $R^2 = 0.85$. This means that as the pore space fractal dimension in the sand packs increased, the pore space fractal dimension in the glass bead packs increased as well (Table 2). Similar results were observed for $h_{min}$ values in both pack types. However, we found a smaller correlation coefficient ($R^2 = 0.56$). We also compared $\beta$ values in the sand packs with those in the glass bead packs and found $R^2 = 0.08$. However, after removing samples SC4 and GC4 from our analysis, we found $(\beta)_G = 2.515(\beta)_S - 0.65$ with $R^2 = 0.73$. We should also note that no significant correlation was found between the residual water contents in the sand and glass bead packs.

### 4.2 Analysis of PSDs and Particle Shape Effects on SHC

#### 4.2.1 Effect of Particle-Size Distribution on SHC

Subclasses A1 and A2 include coarse- and medium-textured media with the narrowest PSDs and least heterogeneous pore structures. Well-sorted samples SA1, SA2, GA1 and GA2 are among the most hydraulically conductive packs in this study (see Table 1). Arya et al. (1999), Tripp (2016),

Won et al. (2019), and many others demonstrated that, generally speaking, as texture becomes finer, PSD becomes broader, pore space would be more heterogeneous, and, consequently, SHC decreases. For example, Won et al. (2019) demonstrated that as the median grain size (d50) decreased, SHC decreased as well. In subclasses A1 and A2, we found the SHC values in the glass bead samples greater than those in the sand packs with the same GMD. These discrepancies can be justified by comparing their pore structures and more specifically, their different critical pore radiui. The value of r$_c$ depends on not only pore structure but also grain arrangements (Ghanbarian, 2020). As reported in Table 2, we found the r$_c$ values in the samples SA1 and SA2, 2 or 3 times smaller than those in the samples GA1 and GA2. Since $SHC \propto r_c^2$ (Katz and Thompson, 1986; 1987) one should expect the measured SHC values in the samples GA1 and GA2 to be nearly one order of magnitude greater than those in the samples SA1 and SA2 (see Table 1), although they have similar GMD values. Our results demonstrate that the SHC value is more influenced by pore structure than grain sizes, even in homogeneous packs studied here. As we showed in Figs. 9 and 10, although the GMD might be an appropriate length scale to characterize the SHC in packs with very narrow particle size distributions, it is not representative in relatively homogeneous and heterogeneous media in which water flow is strongly controlled by geometrical and topological characteristics of pore space, such as pore size distribution, pore shape, surface area, and pore connectivity (Ghanbarian et al., 2017b).

### 4.2.2 Evaluation of CPA and Kozeny-Carman Models

In this section, we compare the SHC estimations via both the CPA approach and the Kozeny-Carman model for the glass bead and sand packs studied here. Results are presented in Figs. 11 and 12 in which the dashed red line denotes the 1:1 line (y = x) and the dotted red lines represent the factor of 4 confidence intervals (y = 0.25x and y = 4x). Using the CPA, we found RMSLE =

0.265 and 0.215 respectively for the sand and the glass bead samples (Fig. 11). However, the RMSLE values were 0.918 and 1.06 the using Kozeny- Carman equation (Fig. 12). We found more accurate estimations via the CPA because this approach captures the pore space heterogeneity, while the Kozeny-Carman model focuses on solid matrix properties, such as the average grain diameter. Our results are in accord with those from Rosas et al. (2014) who reported that Carman-Kozeny model estimations were poorly correlated to the measured ones with errors as large as 500% and more. They examined 431 samples (0.05 mm < d10 < 0.83 mm, 0.09 mm < d60 < 4.29 mm, 1.3 < Cu < 18.3) from different depositional conditions with hydraulic conductivity measured via the constant water head method.

Fig. 11

Fig. 12

Our results indicate that the model of Ghanbarian et al. (2017b) based on the concepts of CPA could estimate the SHC in both homogeneous and relatively heterogeneous sand and glass bead packs reasonably well. From Fig. 11, it can be deduced that the Ghanbarian et al. (2017b) model, Eq. (3), has slight tendency to underestimate the SHC in the sand packs. However, based on the RMSLE values, Eq. (3) estimated the SHC more accurately in glass bead packs. In the sand packs, the calculated RMSLE value was 0.132 for class A, 0.309 for class B and 0.353 for classes C and D. Results showed that as the PSD became broader, the CPA estimations became less accurate, in contrast to the results of Ghanbarian et al. (2017b). This, however, need further investigations using a wider range of packs. In the glass bead packs, the RMSLE values were 0.225 for class A, 0.093 for class B and 0.264 for classes C and D.

The RMSLE values obtained here are comparable with RMSLE = 0.315 reported by Ghanbarian et al. (2017b) who evaluated the critical path analysis approach using average SHC values for 11 soil texture classes and found estimated SHC values within the factor of three confidence level of measured ones.

### 4.2.3 Particle Shape Effect on SHC

Results of paired t-test showed no significant difference between the SHC values for sand and glass bead packs. More specifically, we found that for 8 sand samples (with angular particles), the SHC value was less than that in glass bead packs (with rounded grains). Our results revealed that for the samples in classes B to D, the SHC value fluctuates within samples, and there is no monotonic trend in the data. Using coupled discrete element and pore-scale network modeling, Mahmoodlu et al. (2016) reported that mixing sands with finer particles caused a reduction in the value of permeability, whereas mixing with coarser grains increased the permeability value. They argued that such a tendency is due to the role of finer particles that fill out voids among coarser grains.

Several studies in the literature have addressed the effect of particle shape on SHC by means of numerical simulations (see, e.g., Garcia et al., 2009; Torskaya et al., 2014; Katagiri et al., 2015; Liu and Jeng, 2019; Rokhforouz and Amiri, 2019; Xiong et al., 2019). For example, Garcia et al. (2009) simulated SHC in several granular packs composed of grains polydisperse in terms of particle shape and size. Their results, however, showed that grain shape has a small but non-trivial impact on the SHC. More specifically, they found that sphere packs had SHC 1.6 to 1.8 times greater than other packs composed of irregular grains (see their Fig. 10).

To address the influence of particle shape and, more specifically, aspect ratio on SHC, Katagiri et al. (2015) generated grain packs using the discrete element method (DEM), simulated flow using OpenFOAM, and investigated the effect of aspect ratio on the permeability. They found that the grain aspect ratio affected the void ratio and surface area in their packs. As a consequence, one should expect that the aspect ratio should have some impact on the SHC. Although Katagiri et al. (2015) demonstrated that as the grain aspect ratio varied, the SHC changed as well (see their Fig. 4), those authors did not find any clear trend between aspect ratio and SHC, similar to our experimental results. Torskaya et al. (2014) used pore-scale simulation to study the effect of particle shape on several hydraulic properties, including SHC using four different models. In their first model, particles followed exactly the actual particle shapes and size distribution determined from micro-CT images. In their second model, grain packs were constructed of spherical particles with volumes equal to the actual ones on a grain-by-grain basis. Their third model was similar to their second one. However, they used the surface-to-volume ratio instead of particle volume. Torskaya et al. (2014) constructed their fourth model using grain size distribution of ellipsoids followed the actual one with the same volume and surface-to-volume ratio based on a grain-by-grain analysis. Those authors found the best match between simulations and experiments in grain packs with actual angular particles. Torskaya et al. (2014) found that all models except their first model, constructed based on actual particle shapes, overestimated the SHC (see their Fig. 13). They also stated that "Kozeny–Carman's predictions agree with modeled permeability [SHC] for spherical grain packs but overestimate permeability [SHC] for micro-CT images and non-spherical grain packs when volume-based radii are used to calculate the average grain size in a pack." Their simulations demonstrated that the surface-to-volume ratio and particle shape were fundamental parameters that controlled flow and transport in media with similar porosities. Hamamoto et al. (2016) used micro-CT imaging to study the pore structure and hydraulic properties in sand and glass bead packs of different particle size distributions and grain shapes. Based on their experimental observations, packs constructed of well-rounded sand particles and/or glass beads showed greater pores, higher pore coordination numbers, lower volumetric surface areas, and consequently greater SHC values than those composed of angular grains. They concluded that "The X-ray CT measurements revealed that round sands create more continuous and larger/wider pore networks than angular sands when packed to a similar total porosity, resulting in higher saturated hydraulic conductivity, and higher gas diffusivity and air permeability under relatively dry conditions."

More recently, Liu and Jeng (2019) addressed the effect of several factors such as porosity, particle shape, surface properties, and particle size distribution on SHC. They used the Cellular Automata random growth model and spherical harmonic function to generate particles of various shapes and surface characteristics. The 3D lattice-Boltzmann method was applied to simulate the SHC in grain packs constructed by a random packing algorithm. Those authors claimed that the effect of particle shape and surface characteristics on SHC was non-trivial, similar to the results of Torskaya et al. (2014). Liu and Jeng (2019) also carried out a univariate analysis to study the sensitivity of such factors on the SHC and found that the impact of porosity was greater than that of particle size distribution, particle surface, and particle shape. They demonstrated that under constant porosity and other factors conditions, as particle sphericity increases, the SHC increases as well. Liu and Jeng (2019) stated that "In addition to the influence of particle shape on single-phase flow in porous media, its impact on two-phase flow has also been addressed in the literature. For example, Rokhforouz and Amiri (2019) numerically simulated two-phase flow using the finite-element method under various capillary number and wettability conditions. To address the effect of grain shape, those authors simplified a real pore network to two approximations in which actual grain shapes were (1) replaced with rounded circles, and (2) modified by less than 10%. Although porosity and SHC values did not vary within their three models, Rokhforouz and Amiri (2019) showed that simplifying grain shape affected morphologies of the formed fingers and the volume of trapped oil. Furthermore, they indicated that slight modification of particle sizes caused various displacement behavior for two-phase flow.

## 5. Conclusion

In this study, the effect of particle shape and particle size distribution (PSD) on the fractal properties as well as saturated hydraulic conductivity (SHC) of homogeneous and relatively heterogeneous packs was investigated. For this purpose, 16 sand and 16 glass bead packs with similar particle size distributions were prepared within four classes from A (well-sorted) to D (poorly-sorted). Although measuring the moisture characteristic curve (MCC) using the sandbox and pressure plate methods required accurate characterizations, the main limitation in terms of experimental measurements was measuring the MCCs for well and poorly sorted packs separatedly. Such a procedure was time-consuming and required reaching the equilibrium conditions. In addition to that, preparing sand and glass bead packs with similar porosities was needed accurate packing processes. For this purpose, the studied columns were filled uniformly using the dry packing method. Same samples were used for both MCC and SHC measuremenets. Reaching the steady-state conditions for constant head and SHC measurements also limited the experiments. We used de-aired water during the SHC experiment to reduce the outgassing of dissolved gases into the packs. Results showed that the poorly-sorted samples had more complex pore structures and typically greater pore space fractal dimensions ($D_{MCC}$). We found that $D_{MCC}$ ranged between 0.8 and 2.86 for sand and between -0.18 and 2.81 for glass bead packs. Our results showed that the pore space fractal dimension can take a negative value. We also found that the angular sand packs had generally greater pore space fractal dimensions than the glass bead packs with the same geometric mean diameter (GMD). However, no clear trend was observed between the $D_{MCC}$ and GMD in the studied packs. Interestingly, we found that the $D_{MCC}$ for the sand samples was linearly correlated to that for the glass bead samples with $R^2 = 0.85$ (results not shown). Nearly half of the sand samples composed of angular particles had lower SHC values in comparison to the glass beads constructed of rounded particles. We also found that the critical path analysis model of Ghanbarian et al. (2017b) estimated the SHC in homogeneous and relatively heterogeneous packs reasonably well. More specifically, results showed that the SHC estimations for both sand and glass bead packs were within a factor of four of the measurements on average.

Notation

The following symbols and abbreviations are used in this paper:

Table 3

**Acknowledgements**

ER and KZ are grateful to the Department of Water Engineering, Urmia Lake Research Institute, Urmia University, Iran, where all experiments were conducted, for providing appropriate lab facilities. BG acknowledges Kansas State University for supports through faculty startup funds.

# References

Arasan, S., Akbulut, S., Hasiloglu, A.S., 2011. The relationship between the fractal dimension and shape properties of particles. KSCE J. Civ. Eng. 15, 1219–1225. https://doi.org/10.1007/s12205-011-1310-x

Araujo, G.S., Bicalho, K. V, Tristao, F.A., 2017. Use of digital image analysis combined with fractal theory to determine particle morphology and surface texture of quartz sands. J. Rock Mech. Geotech. Eng. 9, 1131–1139.

Arya, L.M., Leij, F.J., Shouse, P.J., van Genuchten, M.T., 1999. Relationship between the hydraulic conductivity function and the particle-size distribution. Soil Sci. Soc. Am. J. 63, 1063–1070. https://doi.org/10.2136/sssaj1999.6351063x

ASTM:D421−85, 2007. Standard Practice for Dry Preparation of Soil Samples for Particle-Size Analysis and Determination of Soil Constants. ASTM Stand. Int. 85, 1–2. https://doi.org/10.1520/D0421-85R07

ASTM:D6836-16, 2016. Standard test methods for determination of the soil water characteristic curve for desorption using a hanging column, pressure extractor, chilled mirror hygrometer, and/or centrifuge. ASTM Stand. Int. 16. https://doi.org/10.1520/D6836-16

Avnir, D., Biham, O., Lidar, D., Malcai, O., 1998. Is the geometry of nature fractal? Science (80-. ). 279, 39–40. https://doi.org/10.1126/science.279.5347.39

Bird, N.R.A., Perrier, E., Rieu, M., 2000. The water retention function for a model of soil structure with pore and solid fractal distributions. Eur. J. Soil Sci. 51, 55–63. https://doi.org/10.1046/j.1365-2389.2000.00278.x

Brooks, R., Corey, T., 1964. Hydraulic Properties of Porous Media. Hydrol. Pap. Color. State Univ. 24, 37.

Carman, P.C., 1937. Fluid flow through granular beds. Trans. Inst. Chem. Eng. 15, 150–166.

Crawford, J.W., Matsui, N., Young, I.M., 1995. The relation between the moisture-release curve and the structure of soil. Eur. J. Soil Sci. 46, 369–375. https://doi.org/10.1111/j.1365-2389.1995.tb01333.x

Cresswell, H.P., Green, T.W., McKenzie, N.J., 2008. The adequacy of pressure plate apparatus for determining soil water retention. Soil Sci. Soc. Am. J. 72, 41–49. https://doi.org/10.2136/sssaj2006.0182

Dagan, G., Fiori, A., Jankovic, I., 2013. Upscaling of flow in heterogeneous porous formations: Critical examination and issues of principle. Adv. Water Resour. 51, 67–85. https://doi.org/10.1016/j.advwatres.2011.12.017

Dathe, A., Thullner, M., 2005. The relationship between fractal properties of solid matrix and pore space in porous media. Geoderma 129, 279–290. https://doi.org/10.1016/j.geoderma.2005.01.003

Doyen, P.M., 1988. Permeability, conductivity, and pore geometry of sandstone. J. Geophys. Res. Solid Earth 93, 7729–7740. https://doi.org/10.1029/JB093iB07p07729

Dullien, F.A.L., 1992. Porous media: Fluid transport and pore structure. Academic Press, San Diego, CA. Porous media Fluid Transp. pore Struct. Acad. Press. San Diego, CA.

Dullien, F.A.L., Zarcone, C., Macdonald, I.F., Collins, A., Bochard, R.D.E., 1989. The effects of surface roughness on the capillary pressure curves and the heights of capillary rise in glass bead packs. J. Colloid Interface Sci. 127, 362–372. https://doi.org/10.1016/0021-9797(89)90042-8

Garcia, X., Akanji, L.T., Blunt, M.J., Matthai, S.K., Latham, J.P., 2009. Numerical study of the effects of particle shape and polydispersity on permeability. Phys. Rev. E 80, 21304.

https://doi.org/10.1103/PhysRevE.80.021304

Ghanbarian-Alavijeh, B., Hunt, A.G., 2012a. Estimation of soil-water retention from particle-size distribution: Fractal approaches. Soil Sci. 177, 321–326. https://doi.org/10.1097/SS.0b013e3182499910

Ghanbarian-Alavijeh, B., Hunt, A.G., 2012b. Comments on "More general capillary pressure and relative permeability models from fractal geometry" by Kewen Li. J. Contam. Hydrol. 140, 21–23. https://doi.org/10.1016/j.jconhyd.2012.08.004

Ghanbarian, B., 2020. Applications of critical path analysis to uniform grain packings with narrow conductance distributions: I. Single-phase permeability. Adv. Water Resour. 137, 103529. https://doi.org/10.1016/j.advwatres.2020.103529

Ghanbarian, B., Hamamoto, S., Kawamoto, K., Sakaki, T., Moldrup, P., Nishimura, T., Komatsu, T., 2018. Saturation-dependent gas transport in sand packs: Experiments and theoretical applications. Adv. Water Resour. 122, 139–147. https://doi.org/10.1016/j.advwatres.2018.10.011

Ghanbarian, B., Hunt, A.G., Skaggs, T.H., Jarvis, N., 2017a. Upscaling soil saturated hydraulic conductivity from pore throat characteristics. Adv. Water Resour. 104, 105–113. https://doi.org/10.1016/j.advwatres.2017.03.016

Ghanbarian, B., Ioannidis, M.A., Hunt, A.G., 2017b. Theoretical insight into the empirical tortuosity-connectivity factor in the Burdine-Brooks-Corey water relative permeability model. Water Resour. Res. 53, 10395–10410. https://doi.org/10.1002/2017WR021753

Ghanbarian, B., Millán, H., 2017. Fractal Capillary Pressure Curve Models, in: Fractals. CRC Press, pp. 29–54.

Ghanbarian, B., Sahimi, M., 2017. Electrical conductivity of partially saturated packings of

particles. Transp. Porous Media 118, 1–16. https://doi.org/10.1007/s11242-017-0821-4

Ghanbarian, B., Sahimi, M., Daigle, H., 2016. Modeling relative permeability of water in soil: Application of effective-medium approximation and percolation theory. Water Resour. Res. 52, 5025–5040. https://doi.org/10.1002/2015WR017903

Hamamoto, S., Moldrup, P., Kawamoto, K., Sakaki, T., Nishimura, T., Komatsu, T., 2016. Pore network structure linked by X-ray CT to particle characteristics and transport parameters. Soils Found. 56, 676–690. https://doi.org/10.1016/j.sandf.2016.07.008

Hunt, A., Ewing, R., Ghanbarian, B., 2014. Fractal Models of Porous Media, in: Percolation Theory for Flow in Porous Media. Springer, pp. 103–129. https://doi.org/10.1007/978-3-319-03771-4_4

Jačka, L., Pavlásek, J., Kuráž, V., Pech, P., 2014. A comparison of three measuring methods for estimating the saturated hydraulic conductivity in the shallow subsurface layer of mountain podzols. Geoderma 219–220, 82–88. https://doi.org/10.1016/j.geoderma.2013.12.027

Katagiri, J., Saomoto, H., Utsuno, M., 2015. Quantitative evaluation of the effect of grain aspect ratio on permeability. Vadose Zo. J. 14. https://doi.org/10.2136/vzj2014.10.0138

Katz, A.J., Thompson, A.H., 1987. Prediction of rock electrical conductivity from mercury injection measurements. J. Geophys. Res. Solid Earth 92, 599–607. https://doi.org/10.1029/JB092iB01p00599

Katz, A.J., Thompson, A.H., 1986. Quantitative prediction of permeability in porous rock. Phys. Rev. B 34, 8179. https://doi.org/10.1103/PhysRevB.34.8179

Klute, A., Dirksen, C., 1986. Hydraulic conductivity and diffusivity: Laboratory methods. Methods Soil Anal. Part 1 Phys. Mineral. Methods 5, 687–734. https://doi.org/10.2136/sssabookser5.1.2ed.c28

Kozeny, J., 1927. Uber kapillare leitung der wasser in boden. R. Acad. Sci. Vienna, Proc. Cl. I 136, 271–306.

Lewis, J., Sjöstrom, J., 2010. Optimizing the experimental design of soil columns in saturated and unsaturated transport experiments. J. Contam. Hydrol. 115, 1–13. https://doi.org/10.1016/j.jconhyd.2010.04.001

Li, K., Horne, R.N., 2009. Experimental study and fractal analysis of heterogeneity in naturally fractured rocks. Transp. Porous Media 78, 217–231. https://doi.org/10.1007/s11242-008-9295-8

Liu, Y.F., Jeng, D.S., 2019. Pore Scale Study of the Influence of Particle Geometry on Soil Permeability. Adv. Water Resour. https://doi.org/10.1016/j.advwatres.2019.05.024

Mahmoodlu, M.G., Raoof, A., Sweijen, T., Van Genuchten, M.T., 2016. Effects of sand compaction and mixing on pore structure and the unsaturated soil hydraulic properties. Vadose Zo. J. 15. https://doi.org/10.2136/vzj2015.10.0136

Mandelbrot, B.B., 1982. The fractal geometry of nature WH Freeman and Company New York. NY zbMATH.

Mandelbrot, B.B., 1990. Negative fractal dimensions and multifractals. Physica A: Statistical Mechanics and its Applications, 163, 306-315. https://doi.org/10.1016/0378-4371(90)90339-T

Moslehi, M., de Barros, F.P.J., Ebrahimi, F., Sahimi, M., 2016. Upscaling of solute transport in disordered porous media by wavelet transformations. Adv. Water Resour. 96, 180–189. https://doi.org/10.1016/j.advwatres.2016.07.013

Nimmo, J.R., Perkins, K.S., 2002. 2.6 Aggregate stability and size distribution. Methods soil Anal. part 4, 317–328. https://doi.org/10.2136/sssabookser5.4.c14

NRCS, U., 1993. Soil survey division staff (1993) soil survey manual. Soil conservation service. US Dep. Agric. Handb. 18, 315.

O'Carroll, D.M., Bradford, S.A., Abriola, L.M., 2004. Infiltration of PCE in a system containing spatial wettability variations. J. Contam. Hydrol. 73, 39–63. https://doi.org/10.2136/sssaj1996.03615995006000010010x

Oliviera, I.B., Demond, A.H., Salehzadeh, A., 1996. Packing of sands for the production of homogeneous porous media. Soil Sci. Soc. Am. J. 60, 49–53. https://doi.org/10.2136/sssaj1996.03615995006000010010x

Plummer, M.A., Hull, L.C., Fox, D.T., 2004. Transport of carbon-14 in a large unsaturated soil column. Vadose Zo. J. 3, 109–121. https://doi.org/10.2136/vzj2004.1090

Porter, L.B., Ritzi, R.W., Mastera, L.J., Dominic, D.F., Ghanbarian-Alavijeh, B., 2013. The Kozeny-Carman Equation with a Percolation Threshold. Groundwater 51, 92–99. https://doi.org/10.1111/j.1745-6584.2012.00930.x

Rawls, W.J., Brakensiek, D.L., Saxtonn, K.E., 1982. Estimation of soil water properties. Trans. ASAE 25, 1316–1320. https://doi.org/10.1016/j.advwatres.2018.12.008

Rokhforouz, M.R., Amiri, H.A.A., 2019. Effects of grain size and shape distribution on pore-scale numerical simulation of two-phase flow in a heterogeneous porous medium. Adv. Water Resour. 124, 84–95. https://doi.org/10.1016/j.advwatres.2018.12.008

Rosas, J., Lopez, O., Missimer, T.M., Coulibaly, K.M., Dehwah, A.H.A., Lujan, L.R., Mantilla, D., 2014. Determination of Hydraulic Conductivity from Grain-Size Distribution for Different Depositional Environments 52, 399–413. https://doi.org/10.1111/gwat.12078

Sahimi, M., 2003. Heterogeneous Materials: Nonlinear and breakdown properties and atomistic modeling. Springer Science & Business Media.

Soldi, M., Guarracino, L., Jougnot, D., 2017. A simple hysteretic constitutive model for unsaturated flow. Transp. Porous Media 120, 271–285. https://doi.org/10.1007/s11242-017-0920-2

Sperry, J.M., Peirce, J.J., 1995. A model for estimating the hydraulic conductivity of granular material based on grain shape, grain size, and porosity. Groundwater 33, 892–898. https://doi.org/10.1111/j.1745-6584.1995.tb00033.x

Stakman, W.P., Valk, G.A., Van der Harst, G.G., 1969. Determination of soil moisture retention curves. I. Sand box Appar. Range pF 0 to 2, 1–19.

Taylor, K.E., 2001. Summarizing multiple aspects of model performance in a single diagram. J. Geophys. Res. Atmos. 106, 7183–7192. https://doi.org/10.1029/2000JD900719

Torskaya, T., Shabro, V., Torres-Verdín, C., Salazar-Tio, R., Revil, A., 2014. Grain shape effects on permeability, formation factor, and capillary pressure from pore-scale modeling. Transp. Porous Media 102, 71–90. https://doi.org/10.1007/s11242-013-0262-7

Tripp, B.J., 2016. Dependence of transport properties on grain size distribution.

Won, J., Park, J., Choo, H., Burns, S., 2019. Estimation of saturated hydraulic conductivity of coarse-grained soils using particle shape and electrical resistivity. J. Appl. Geophys. 167, 19–25. https://doi.org/10.1016/j.jappgeo.2019.05.013

Xiong, Y., Long, X., Huang, G., Furman, A., 2019. Impact of pore structure and morphology on flow and transport characteristics in randomly repacked grains with different angularities. Soils Found. https://doi.org/10.1016/j.sandf.2019.10.002

Xu, P., Yu, B., 2008. Developing a new form of permeability and Kozeny–Carman constant for homogeneous porous media by means of fractal geometry. Adv. Water Resour. 31, 74–81. https://doi.org/10.1016/j.advwatres.2007.06.003
31

# Table Captions

**Table 1.** The salient properties of the samples studied.

**Table 2**. The optimized parameters for sand and glass bead packs.

**Table 3**. The list of symbles and abbriviations.

**Table 1** The salient properties of the samples studied

| No. | Class name | Subclass name | Mesh Size | GMD (mm) | Diameter Classification (Soil Survey Division Staff, 1993) | Sand Packs | | | | Glass Bead Packs | | | |
|---|---|---|---|---|---|---|---|---|---|---|---|---|---|
| | | | | | | Sample Name | $\rho_b$ (g/cm$^3$) | $\phi$ (cm$^3$/cm$^3$) | Saturated Hydraulic Conductivity (cm/h) | Sample Name | $\rho_b$ (g/cm$^3$) | $\phi$ (cm$^3$/cm$^3$) | Saturated Hydraulic Conductivity (cm/h) |
| 1 | A | A1 | 30-25* | 0.65 | Coarse | SA1 | 1.55 | 0.42 | 308.83 | GA1 | 1.61 | 0.39 | 2405.06 |
| 2 | | A2 | 40-30 | 0.50 | Medium | SA2 | 1.55 | 0.41 | 205.20 | GA2 | 1.66 | 0.37 | 2363.19 |
| 3 | | A3 | 50-40 | 0.32 | Medium | SA3 | 1.55 | 0.41 | 61.06 | GA3 | 1.61 | 0.39 | 47.58 |
| 4 | | A4 | 70-50 | 0.25 | Fine | SA4 | 1.56 | 0.41 | 20.40 | GA4 | 1.59 | 0.40 | 133.95 |
| 5 | | A5 | 100-70 | 0.16 | Fine | SA5 | 1.54 | 0.42 | 28.45 | GA5 | 1.58 | 0.40 | 14.61 |
| 6 | | A6 | 120-100 | 0.14 | Fine | SA6 | 1.53 | 0.42 | 11.37 | GA6 | 1.53 | 0.42 | 11.74 |
| 7 | | A7 | 140-120 | 0.11 | Very Fine | SA7 | 1.58 | 0.41 | 9.06 | GA7 | 1.58 | 0.42 | 7.07 |
| 8 | B | B1** | A1+ A2+A3 | 0.48 | Medium | SB1 | 1.53 | 0.42 | 209.16 | GB1 | 1.66 | 0.37 | 68.34 |
| 9 | | B2 | A5+ A6+ A7 | 0.14 | Fine | SB2 | 1.73 | 0.36 | 2.98 | GB2 | 1.73 | 0.35 | 9.00 |
| 10 | | B3 | A3+A4+A5 | 0.24 | Fine | SB3 | 1.58 | 0.40 | 40.33 | GB3 | 1.58 | 0.40 | 15.07 |
| 11 | | B4 | A1+ A4+ A7 | 0.27 | Medium | SB4 | 1.56 | 0.41 | 24.01 | GB4 | 1.56 | 0.41 | 1.01 |
| 12 | C | C1 | A1+ A2+A3+A4+A5 | 0.34 | Medium | SC1 | 1.57 | 0.41 | 48.32 | GC1 | 1.57 | 0.41 | 56.26 |
| 13 | | C2 | A3+A4+A5+ A6+ A7 | 0.18 | Fine | SC2 | 1.58 | 0.40 | 21.81 | GC2 | 1.58 | 0.40 | 6.70 |
| 14 | | C3 | A1+ A2+A4+ A6+ A7 | 0.27 | Medium | SC3 | 1.57 | 0.41 | 14.39 | GC3 | 1.58 | 0.41 | 20.88 |
| 15 | | C4 | A2+A3+A4+A5+ A6 | 0.25 | Fine | SC4 | 1.58 | 0.40 | 29.98 | GC4 | 1.58 | 0.40 | 43.13 |
| 16 | D | | A1+A2+…+ A7 | 0.25 | Fine | SD | 1.58 | 0.40 | 59.87 | GD | 1.58 | 0.40 | 3.85 |

* Sieve No.

**Subclasses B1 to D were formed using equal fraction of each component

**Table 2** The optimized parameters for sand and glass bead packs

| Sample NO. | Pack | Class name | Samples | Eqs. (1) and (2) | | | | | | $r_{max}$ (cm) | $r_c$ (cm) |
|---|---|---|---|---|---|---|---|---|---|---|---|
| | | | | $h_{min}$ (cmH$_2$O) | $D_{MCC}$ | $\beta$ | $\Theta_r$ * (cm$^3$/cm$^3$) | RMSE (cm$^3$/cm$^3$) | $R^2$ | | |
| 1 | | | SA1 | 0.01 | 1.29 | 0.38 | 0.04 | 0.00 | 1.00 | 0.027 | 0.025 |
| 2 | | | SA2 | 0.01 | 1.25 | 0.35 | 0.04 | 0.01 | 0.99 | 0.017 | 0.016 |
| 3 | | | SA3 | 0.01 | 0.80 | 0.37 | 0.05 | 0.00 | 1.00 | 0.010 | 0.009 |
| 4 | | A | SA4 | 0.03 | 1.00 | 0.39 | 0.07 | 0.01 | 1.00 | 0.005 | 0.005 |
| 5 | | | SA5 | 0.03 | 1.51 | 0.45 | 0.07 | 0.01 | 1.00 | 0.006 | 0.005 |
| 6 | | | SA6 | 0.03 | 1.80 | 0.40 | 0.04 | 0.01 | 0.98 | 0.005 | 0.004 |
| 7 | | | SA7 | 0.05 | 1.44 | 0.40 | 0.05 | 0.03 | 0.97 | 0.003 | 0.003 |
| 8 | Sand | | SB1 | 0.01 | 1.90 | 0.36 | 0.05 | 0.01 | 1.00 | 0.017 | 0.015 |
| 9 | | B | SB2 | 0.05 | 1.87 | 0.56 | 0.08 | 0.00 | 0.99 | 0.005 | 0.004 |
| 10 | | | SB3 | 0.03 | 1.94 | 0.46 | 0.05 | 0.01 | 0.99 | 0.006 | 0.005 |
| 11 | | | SB4 | 0.03 | 2.02 | 0.32 | 0.04 | 0.01 | 1.00 | 0.003 | 0.003 |
| 12 | | | SC1 | 0.02 | 1.89 | 0.33 | 0.05 | 0.00 | 1.00 | 0.010 | 0.008 |
| 13 | | C | SC2 | 0.03 | 2.46 | 0.75 | 0.06 | 0.02 | 0.99 | 0.005 | 0.004 |
| 14 | | | SC3 | 0.03 | 2.19 | 0.32 | 0.08 | 0.01 | 1.00 | 0.008 | 0.005 |
| 15 | | | SC4 | 0.02 | 2.86 | 1.74 | 0.08 | 0.00 | 0.99 | 0.005 | 0.004 |
| 16 | | D | SD | 0.03 | 1.75 | 0.37 | 0.07 | 0.02 | 0.98 | 0.006 | 0.005 |
| 17 | | | GA1 | 0.00 | 0.83 | 0.30 | 0.05 | 0.01 | 1.00 | 0.065 | 0.060 |
| 18 | | | GA2 | 0.00 | 1.0 | 0.28 | 0.09 | 0.00 | 1.00 | 0.059 | 0.049 |
| 19 | | | GA3 | 0.01 | -0.18 | 0.28 | 0.06 | 0.00 | 1.00 | 0.010 | 0.009 |
| 20 | | A | GA4 | 0.01 | 0.77 | 0.26 | 0.09 | 0.00 | 1.00 | 0.011 | 0.009 |
| 21 | | | GA5 | 0.03 | 1.39 | 0.26 | 0.10 | 0.01 | 0.97 | 0.005 | 0.004 |
| 22 | | | GA6 | 0.06 | 1.10 | 0.39 | 0.03 | 0.01 | 0.98 | 0.002 | 0.002 |
| 23 | | | GA7 | 0.05 | 1.20 | 0.50 | 0.01 | 0.01 | 1.00 | 0.003 | 0.003 |
| 24 | Glass Beads | | GB1 | 0.01 | 1.90 | 0.32 | 0.05 | 0.01 | 1.00 | 0.015 | 0.013 |
| 25 | | B | GB2 | 0.04 | 1.42 | 0.38 | 0.03 | 0.02 | 0.96 | 0.004 | 0.004 |
| 26 | | | GB3 | 0.02 | 2.07 | 0.26 | 0.05 | 0.01 | 0.99 | 0.007 | 0.005 |
| 27 | | | GB4 | 0.05 | 1.80 | 0.22 | 0.15 | 0.01 | 0.98 | 0.003 | 0.001 |
| 28 | | | GC1 | 0.01 | 1.66 | 0.28 | 0.09 | 0.01 | 1.00 | 0.011 | 0.008 |
| 29 | | C | GC2 | 0.03 | 2.81 | 1.55 | 0.08 | 0.00 | 1.00 | 0.006 | 0.005 |
| 30 | | | GC3 | 0.02 | 2.32 | 0.26 | 0.15 | 0.03 | 0.94 | 0.010 | 0.003 |
| 31 | | | GC4 | 0.01 | 2.57 | 0.35 | 0.08 | 0.01 | 0.99 | 0.016 | 0.009 |
| 32 | | D | GD | 0.03 | 1.81 | 0.26 | 0.15 | 0.01 | 0.99 | 0.005 | 0.003 |

* The residual water content was determined by directly fitting Eq. (2) to the measured MCCs.

**Table 3**. The list of symbles and abbriviations.

| Symbles | Description | Symbles | Description |
|---|---|---|---|
| MCC | moisture characteristic curve | $r_{max}$ | largest pore radii in the medium |
| SHC | saturated hydraulic conductivity | $r_c$ | critical pore radius |
| PSD | particle size distribution | $f_f$ | fluidity factor |
| $D_{MCC}$ | pore space fractal dimension | $\rho_f$ | fluid density |
| CPA | critical path analysis | $g$ | gravitational acceleration |
| $\theta$ | water content | $\mu$ | dynamic fluid viscosity |
| $\theta_t$ | critical water content | $A$ | constant coefficient in the Young-Laplace equation |
| $\theta_r$ | residual water content | $\gamma$ | air-water interfacial tension |
| $\rho_b$ | bulk density | $\omega$ | air-water contact angle |
| $\rho_s$ | particle density | $C_{KT}$ | Katz-Thompson constant coefficient |
| $\phi$ | porosity | $\bar{D}$ | representative grain diameter |
| GMD | geometric mean diameter | $R^2$ | correlation coefficient |
| $\beta$ | fitting parameter of Eq. (1) | RMSE | root mean square error |
| $h$ | suction (or tension) head | RMSLE | root mean square log-transformed error |
| $h_{min}$ | air entry pressure | $P_F^{(i)}$ | fitted values |
| $S_w$ | water saturation | $P_M^{(i)}$ | measured values |
| $S_{wr}$ | residual water saturation | $n$ | number of the data points |
| $r_{min}$ | smallest pore radii in the medium | | |

# Figure Captions

**Fig. 1** Well-rounded spherical glass bead particles (a) versus angular sand grains (b)

**Fig. 2** Variation of GMD and DMCC for sand (a) and glass bead (b) packs in class A

**Fig. 3** The measured MCC and the fitted model, Eq. (1), for samples in class A. The letters M and F respectively represent measured and fitted

**Fig. 4** Variation of GMD and $D_{MCC}$ for sand and glass bead packs in class B

**Fig. 5** The measured MCC and the fitted model, Eq. (1), for samples in class B. The letters M and F respectively represent measured and fitted

**Fig. 6** Variation of GMD and fractal dimention for sand and glass bead packs in classes C and D

**Fig. 7** The measured MCC and the fitted model, Eq. (1), for samples in classes C and D. The letters M and F respectively represent measured and fitted

**Fig. 8** Taylor diagram for all sand (a) and glass bead (b) packs

**Fig. 9** Measured SHC as a function of the GMD values for all sand (a) and glass bead (b) packs

**Fig. 10** Measured SHC as a function of the $r_c$ values for all sand (a) and glass bead (b) packs

**Fig. 11** The calculated saturated hydraulic conductivity (SHC) using the Ghanbarian et al. (2017b) model, Eq. (3), against the measured one in sand (a) and glass bead (b) packs. RMSLE denotes the root mean square log-transformed error. The dashed and dotted lines represent the 1:1 line and factor of four confidence intervals, respectively

**Fig. 12** The calculated saturated hydraulic conductivity (SHC) using the Kozeny-carman model, Eq. (4), against the measured one in sand (a) and glass bead (b) packs. RMSLE denotes the root mean square log-transformed error. The dashed and dotted lines represent the 1:1 line

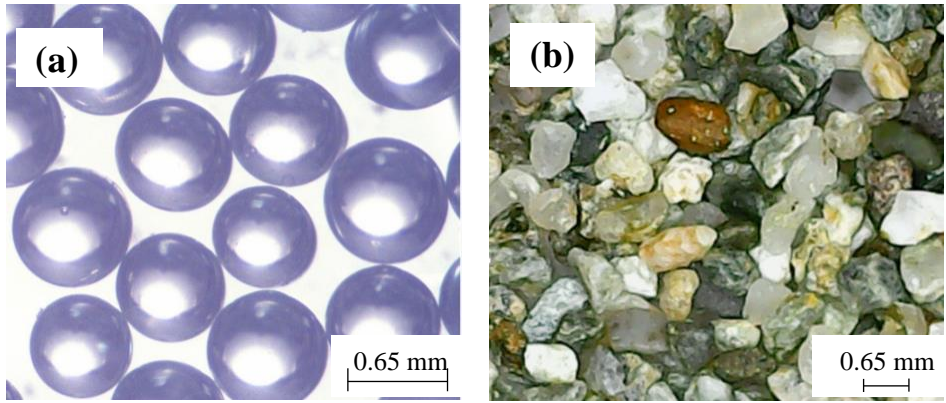

**Fig. 1** Well-rounded spherical glass bead particles (a) versus angular sand grains (b)

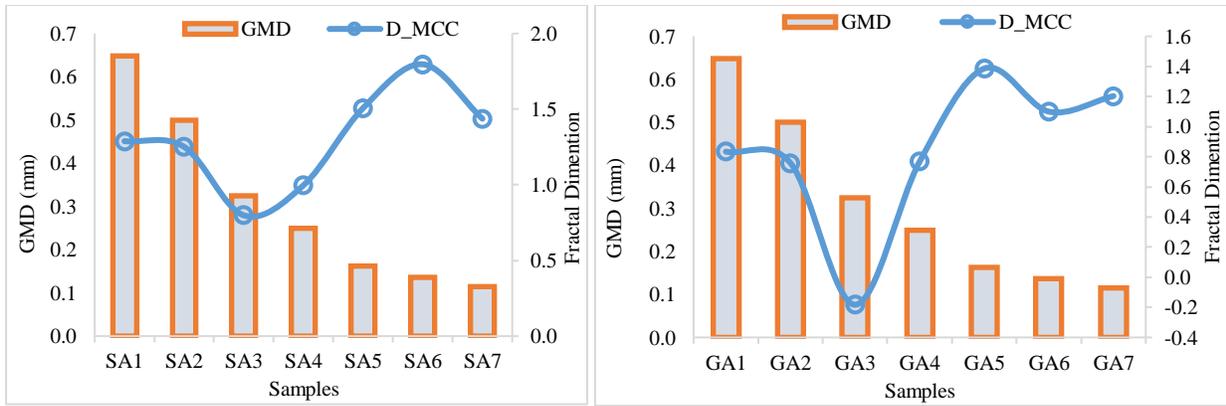

**Fig. 2** Variation of GMD and $D_{MCC}$ for sand (a) and glass bead (b) packs in class A

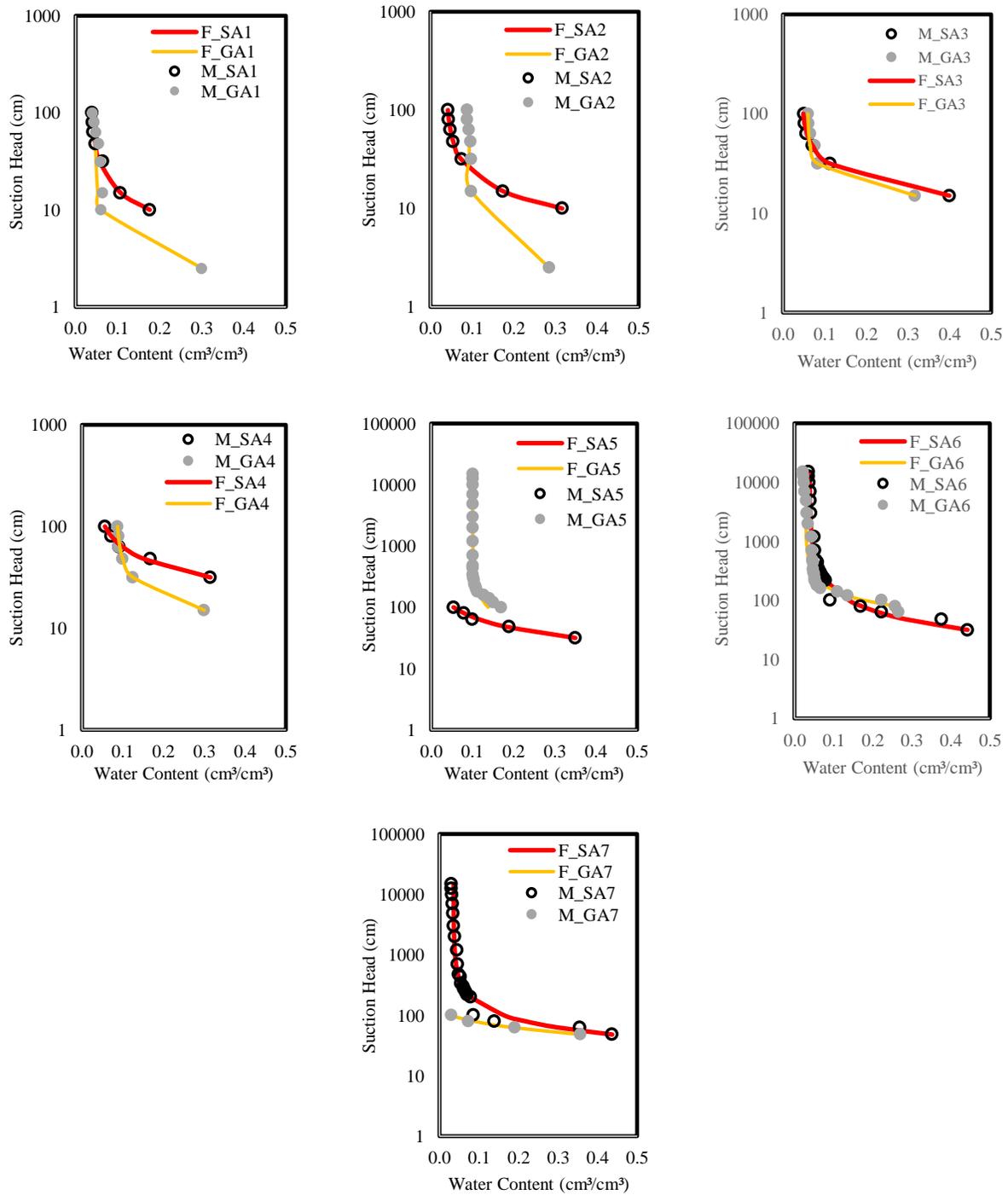

**Fig. 3** The measured MCC and the fitted model, Eq. (1), for samples in class A. The letters M and F respectively represent measured and fitted

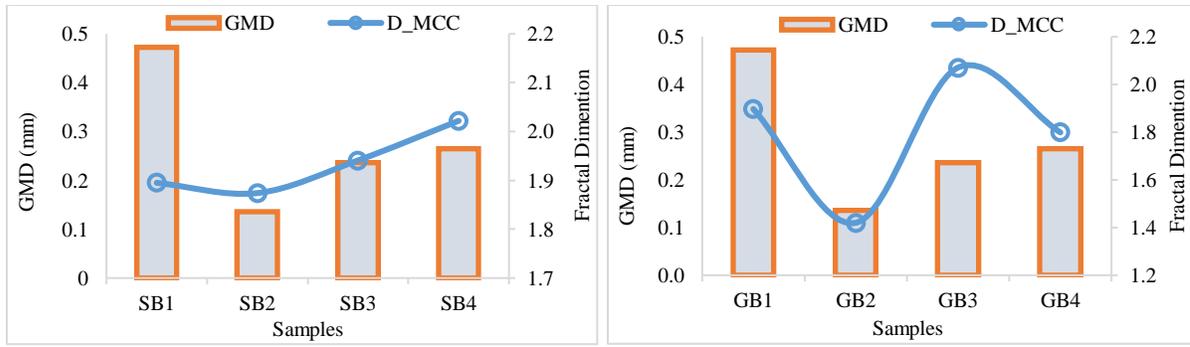

**Fig. 4** Variation of GMD and $D_{MCC}$ for sand and glass bead packs in class B

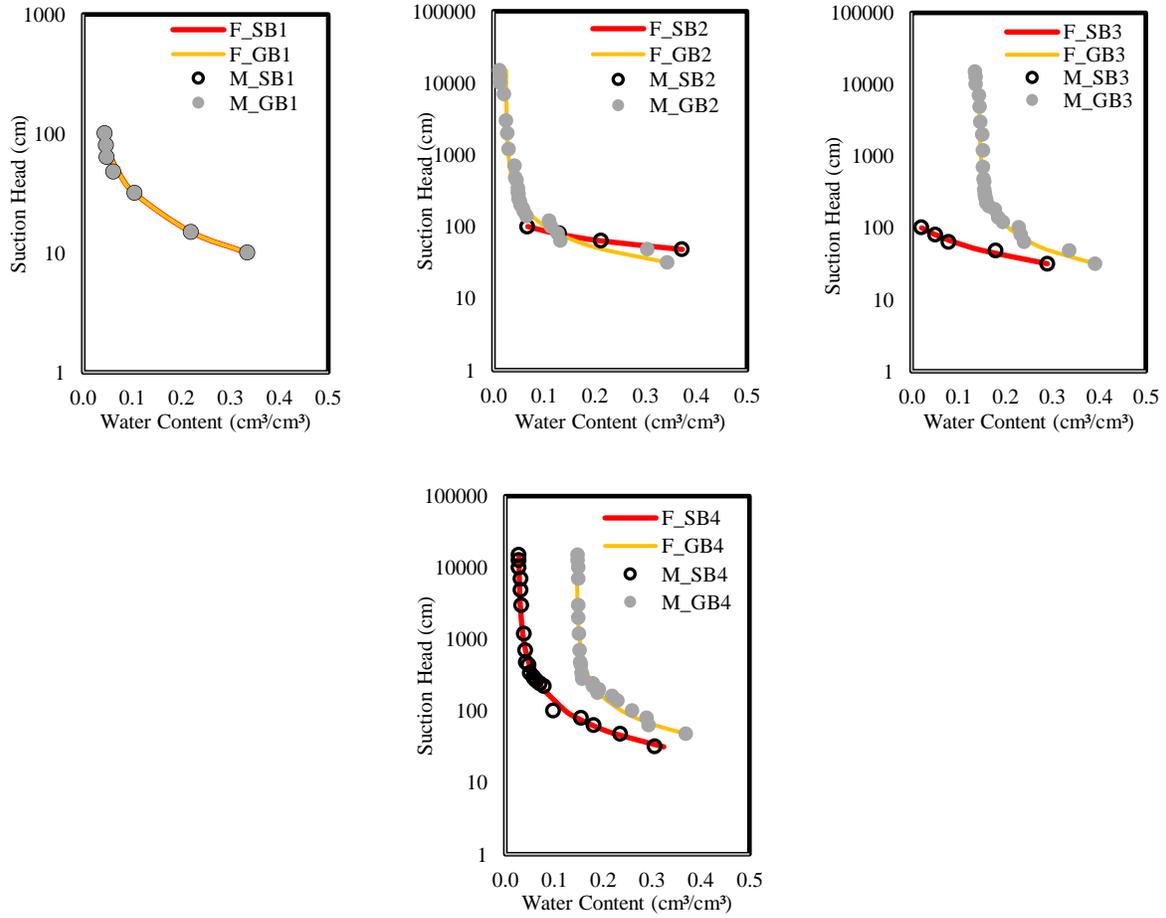

**Fig. 5** The measured MCC and the fitted model, Eq. (1), for samples in class B. The letters M and F respectively represent measured and fitted

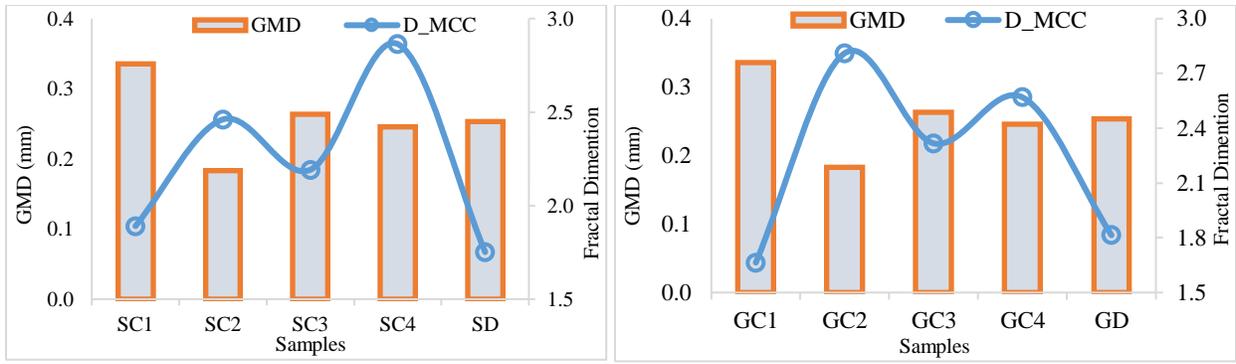

**Fig. 6** Variation of GMD and fractal dimention for sand and glass bead packs in classes C and D

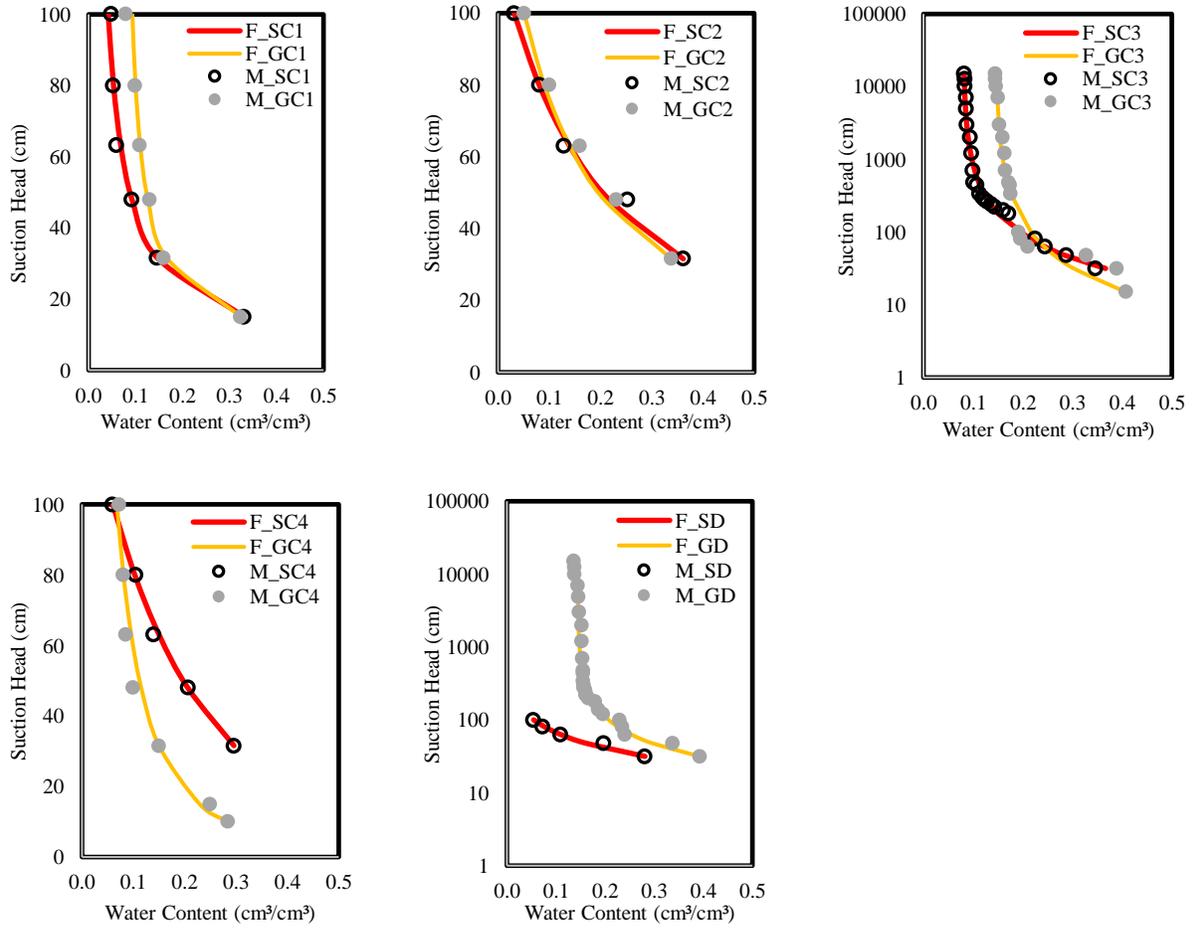

**Fig. 7** The measured MCC and the fitted model, Eq. (1), for samples in classes C and D. The letters M and F respectively represent measured and fitted

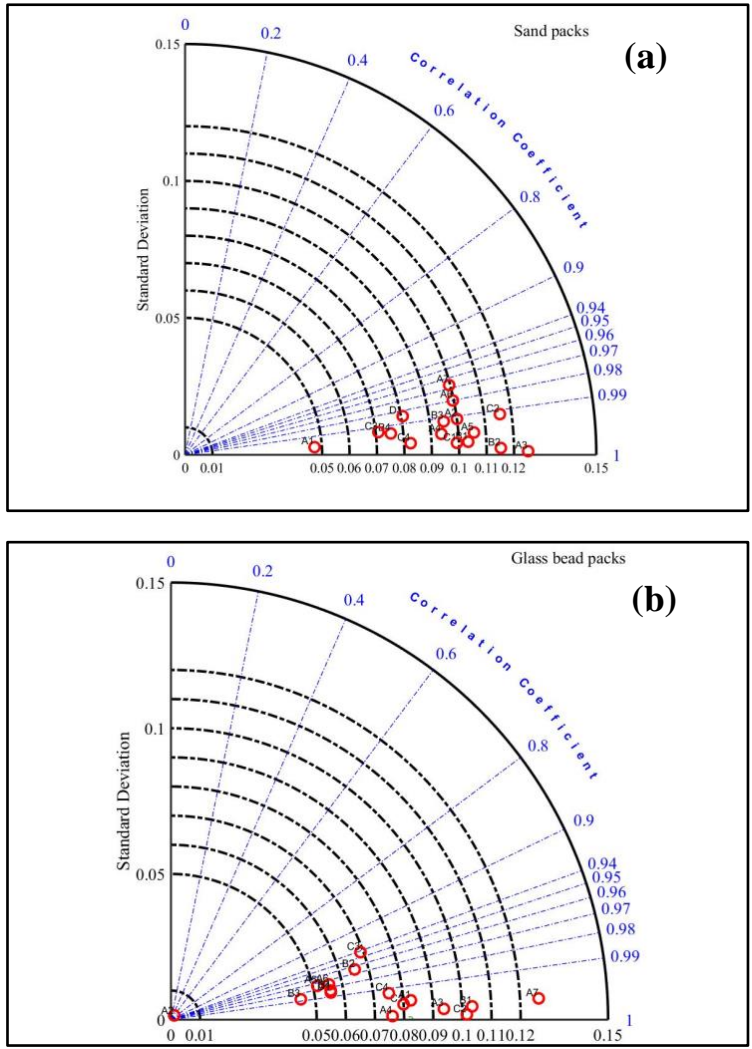

**Fig. 8** Taylor diagram for all sand (a) and glass bead (b) packs

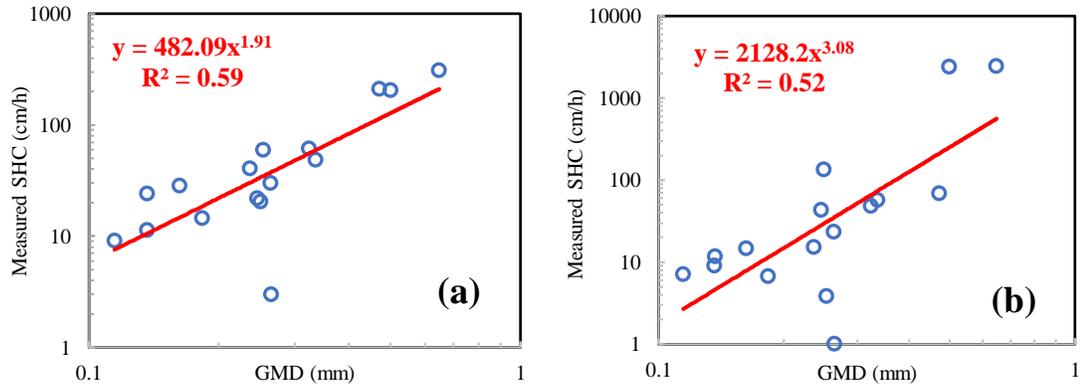

**Fig. 9** Measured SHC as a function of the GMD values for all sand (a) and glass bead (b) packs

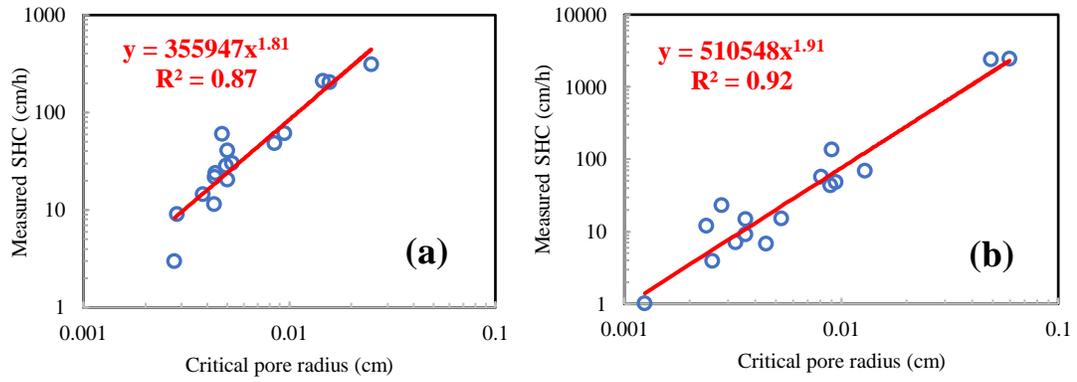

**Fig. 10** Measured SHC as a function of the $r_c$ values for all sand (a) and glass bead (b) packs

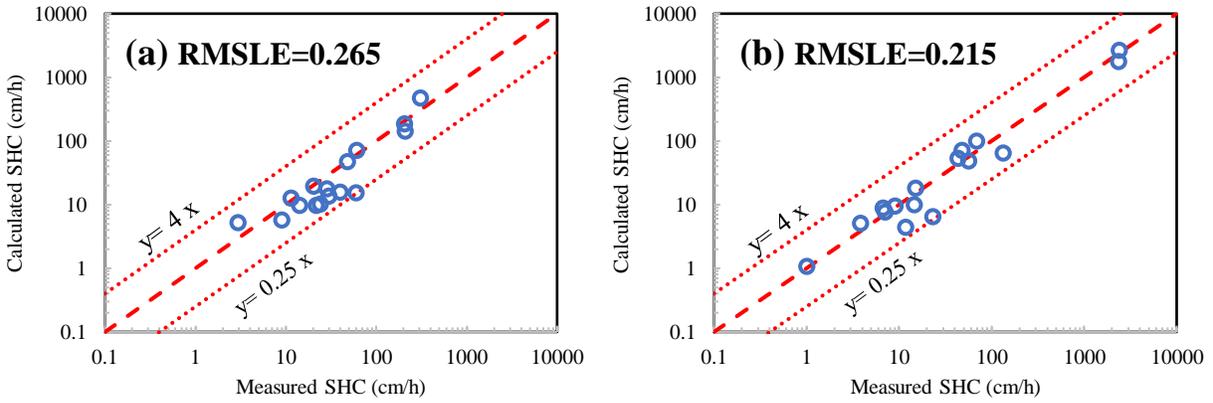

**Fig. 11** The calculated saturated hydraulic conductivity (SHC) using the Ghanbarian et al. (2017b) model, Eq. (3), against the measured one in sand (a) and glass bead (b) packs. RMSLE denotes the root mean square log-transformed error. The dashed and dotted lines represent the 1:1 line and factor of four confidence intervals, respectively

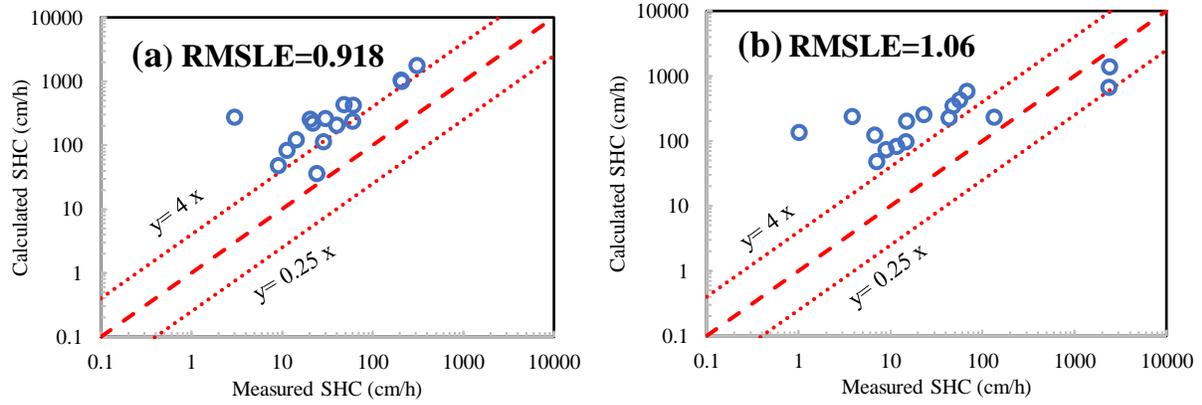

**Fig. 12** The calculated saturated hydraulic conductivity (SHC) using the Kozeny-carman model, Eq. (4), against the measured one in sand (a) and glass bead (b) packs. RMSLE denotes the root mean square log-transformed error. The dashed and dotted lines represent the 1:1 line